\documentclass[%
 reprint,
 amsmath,amssymb,
 aps,
]{revtex4-1}
\usepackage{url}
\usepackage{hyperref}
\usepackage{subfigure}
\usepackage{graphicx}
\usepackage{dcolumn}
\usepackage{bm}

\def\beq{\begin{equation}}
\def\eeq{\end{equation}}
\def\be{\begin{equation}}
\def\ee{\end{equation}}
\def\bea{\begin{eqnarray}}
\def\eea{\end{eqnarray}}

\begin{document}
\title{Stable Cosmic Time Crystals}
\medskip\
\author{ Damien A. Easson}%
\email[Email:]{easson@asu.edu}
\affiliation{Department of Physics,
Arizona State University, Tempe, AZ 85287-1504, USA}
\author{Tucker Manton} 
\email[Email:]{tucker.manton@asu.edu}
\affiliation{Department of Physics,
Arizona State University, Tempe, AZ 85287-1504, USA}

\date{\today}

\begin{abstract}
Cosmological time crystals are created when a scalar field moves periodically through phase space in a spatially flat Friedmann-Robertson-Walker spacetime due to the presence of a limit cycle.
All such cosmological time crystals in the literature suffer from gradient instabilities occurring at Null Energy Condition violating phases where the square sound speed for cosmological perturbations becomes negative. Here we present stable cosmological time crystals. Our analysis suggests this new form of scalar matter--cosmic time crystals--may be considered as a physically viable cosmological matter source.
\end{abstract}

\pacs{Valid PACS appear here}
\maketitle


Dynamical systems can display motion even in their lowest energy states. Considering such systems lead to the creation of classical, and then quantum, \it time crystals\rm~\cite{Wilczek1,Wilczek2}. Recently, an exciting 
field theory extension explored scalar field oscillatory solutions in a cosmological context~\cite{Wilczek3}; however, it was shown that the phase space of such solutions was plagued with regions containing ghost degrees of freedom or imaginary sound speed $c_{s}^{2}<0$, referred to as a gradient instability~\cite{Damien1}. Here we present a method to cure gradient instabilities in such solutions and provide the first examples of stable cosmological time crystals. These novel fields are the first of their kind and have many different applications in cosmology. We organize this note as follows: after a short description of the background and its instability, we briefly describe the effective field theory (EFT) of cosmological perturbations. We then show how the presence of second order operators cures the instability and numerically illustrate the effect on particular time crystal solutions. We conclude with some discussion on specific applications of the fields and comment on phenomenological constraints.

First presented in~\cite{Wilczek3}, the time crystals are obtained from a so-called \it k-essence \rm theory
\begin{equation}\label{Sphi}
    S_{\varphi}=\int d^{4}x\sqrt{-g}p(\varphi, X).
\end{equation}
The Lagrangian $p(\varphi,X)$ is interpreted as an effective pressure. The variable $X$ is defined as $X=-\frac{1}{2}(\nabla^{\mu}\varphi)(\nabla_{\mu}\varphi)$, and is non-negative around any time-like background~\footnote{Throughout we take $M_{p}^{2}=1$ and adopt the mostly plus metric signature, opposite to \cite{Damien1}. Note, $X$ is sometimes defined in the literature without the factor of $1/2$.}. This kinetic term is considered an argument of the Lagrangian, differing from the usual form $\mathcal{L}(\varphi,\nabla\varphi)$. The action (\ref{Sphi}) belongs to the class of models introduced in \cite{Horndeski} which preserve second order equations of motion (see \cite{Langlois1} for a treatment of a larger class of theories that satisfy this property). 

The energy momentum tensor (EMT) associated with (\ref{Sphi}) is $T_{\mu\nu}=p_{X}\nabla_{\mu}\varphi\nabla_{\nu}\varphi+pg_{\mu\nu}$ where $p_{X}=\frac{\partial p}{\partial X}.$ In the frame of the four velocity $u_{\mu}=\frac{\nabla_{\mu}\varphi}{\sqrt{2X}},
$ the EMT takes the form of a perfect fluid $T_{\mu\nu}=(\varepsilon+p)u_{\mu}u_{\nu}+pg_{\mu\nu}$ with energy density $\varepsilon=2Xp_{X}-p.$ We consider only isotropic fields $(\varphi = \varphi(t)$) which obey the the equation of motion
\begin{equation}\label{PhiEom}
    \varepsilon_{X}\ddot{\varphi}+3Hp_{X}\dot{\varphi}+\varepsilon_{\varphi}=0.
\end{equation}
Ghost instabilities occur when the 'mass-term' $\varepsilon_{X}=\frac{\partial\varepsilon}{\partial X}$ in (\ref{PhiEom}) becomes negative. Avoiding the $\varepsilon_{X}<0$ region in the phase space constrains any physical solutions, since the equations of motion are singular when $\varepsilon_{X}=0$. The sound speed, the main quantity of interest in this note, can be computed by taking the ratio
$c_{s}^{2}=\frac{p_{X}}{\varepsilon_{X}}=\frac{\partial p}{\partial \varepsilon}\big|_{\varphi}.$ Note that $c_{s}^{2}$ multiplies the first order derivative term in (\ref{PhiEom}) when $\varepsilon_{X}$ is divided out. If $c_{s}^{2}<0$ the equation of motion clearly becomes elliptical, and short-wavelength quantum fluctuations grow without bound \cite{Mukhanov}. In addition, the Null Energy Condition (NEC) $T_{\mu\nu}n^{\mu}n^{\nu}>0$ is violated whenever $p_{X}<0,$ as
\begin{equation}
    T_{\mu\nu}n^{\mu}n^{\nu}=p_{X}(\nabla_{\mu}\varphi n^{\mu})^{2}=\frac{\varepsilon+p}{2X}(\nabla_{\mu}\varphi n^{\mu})^{2}
\end{equation}
for arbitrary null vectors $n^{\mu}.$ We will show that $\varepsilon_{X}<0$ and $c_{s}^{2}<0$ can be avoided while stably violating the NEC, resulting in a physically plausible model. This is achieved by use of the EFT of cosmological perturbations \cite{Creminelli1,Cai2,Langlois3}, the basics of which we will now briefly outline. 

The effective action is derived via a 3+1 decomposition using the ADM metric
\begin{equation}\label{ADM}
    ds^{2}=-N^{2}dt^{2}+h_{ij}(dx^{i}+N^{i}dt)(dx^{j}+N^{j}dt),
\end{equation}
where $h_{ij}$ lowers the Latin indices $i,j=1,2,3$ corresponding to the spatial coordinates and $N$, $N^{i}$ are the usual lapse function and shift vectors. (See \cite{Poisson} for an excellent treatment of this formalism.) The spacetime is decomposed on constant time hypersurfaces where $\varphi$ is uniform, \it i.e.\rm, $\delta\varphi(\bm{x},t)=0$, and only metric perturbations need to be considered. With this choice of slicing, which is called the unitary gauge, the graviton describes three degrees of freedom. The theory is then built from the lowest dimensional operators invariant under spatial diffeomorphisms (namely the inverse metric component $g^{00}$, and the extrinsic and intrinsic curvature tensors). The resulting Lagrangian describes the metric perturbations about a FRW solution consistent with the spatial diffeomorphism invariance~\footnote{For a detailed derivation of the full theory, see \cite{Creminelli2}. Note that a few different conventions and notations are used in the literature, and we will follow that of \cite{Gleyzes:2013ooa}.}. The general EFT action is of the form
\begin{equation}\label{General}
    S=\int d^{4}\! x\sqrt{-g} \, \mathcal{L}(N,K_{\mu\nu},\mathcal{R}_{\mu\nu};t) \,,
\end{equation}
where $\mathcal{R}_{\mu\nu}$ is the spatial Ricci tensor and $K_{\mu\nu}$ is the extrinsic curvature on the constant time hypersurfaces. We obtain the second order action by expanding the Lagrangian (\ref{General}) and using the definitions $\delta K\equiv K-3H,$ $\delta K_{\mu\nu}\equiv K_{\mu\nu}-Hh_{\mu\nu},$ along with 
\bea
    K_{\mu\nu}K^{\mu\nu} &=& 3H^{2}-\delta (K_{\mu\nu}K^{\mu\nu}), \nonumber \\ 
   \delta (K_{\mu\nu}K^{\mu\nu}) &=& 2H\delta K+\delta K^{\mu}_{\nu}\delta K^{\nu}_{\mu} \, ,
\eea
where $h_{\mu\nu}$ is the induced metric on the hypersurface, $K=K^{\mu}_{\mu}$, and $H\equiv\frac{\dot{a}}{a}$ is the Hubble parameter. (Each variation vanishes at the background level, and the resulting action vanishes at first order in the perturbations.) Explicitly, the EFT Lagrangian linear in the perturbations producing only first order derivatives in the fluctuations is
\bea \label{EFT}
    \mathcal{L}&=&\frac{1}{2}f(t)R-\lambda(t)-c(t)g^{00} \\
    &-& \frac{1}{2}m_{3}^{3}(t)\delta K\delta g^{00}-m_{4}^{2}(t)\Big( (\delta K)^{2}-\delta K^{\mu}_{ \ \nu}\delta K^{\nu}_{ \ \mu}\Big)\nonumber \\
    &+& \frac{1}{2}\tilde{m}_{4}^{2}(t)\mathcal{R}\delta g^{00}+\frac{1}{2}M_{2}^{4}(t)(\delta g^{00})^{2} \nonumber.
\eea
In the above, the first line describes the background and each expansion coefficient has time dependence. On the flat ($k=0$) FRW universe considered here, the intrinsic curvature vanishes, $\mathcal{R}^{\mu}_{\nu}=0.$ Consequently, the combination $\mathcal{R}\delta g^{00}$ is second order. We use the inverse metric component $g^{00}$ in the action with its perturbation defined as $\delta g^{00}\equiv g^{00}+1,$ which is easily interchanged with the lapse function $N$ using $g^{00}=-1/N^{2}.$ A key feature of this approach is that none of the quantities in the second two lines effect the background. This provides some freedom to consistently set coefficients to zero while preserving the time crystal structure of the scalar fields.

Computing the (unitary gauge) action for the scalar metric perturbation~\footnote{Here we only consider the scalar perturbations. For a complete treatment of both scalar and tensor perturbations, see e.g. \cite{Cai2}.} $\zeta$ in the EFT gives us an expression for the sound speed $c_{s}^{2}$ in terms of the EFT operators. This is accomplished by taking the spatial metric to be $h_{ij}=a(t)^{2}e^{2\zeta}\delta_{ij}$ and using 
\begin{equation}
    K_{ij}=\frac{1}{2N}\Big(\dot{h}_{ij}-\nabla_{i}N_{j}-\nabla_{j}N_{i}\Big)
\end{equation}
to compute the spatial components of the extrinsic curvature, where $\nabla_{i}$ is the covariant derivative compatible with $h_{ij}$. The end result is an action of the form
\begin{equation}\label{Szeta}
    S^{(2)}_{\zeta}=\int d^{4}x\frac{a^{3}}{2}\Bigg[\mathcal{L}_{\dot{\zeta}\dot{\zeta}}\dot{\zeta}^{2}+\mathcal{L}_{\partial_{i}\zeta\partial_{i}\zeta}\frac{(\partial_{i}\zeta)^{2}}{a^{2}}\Bigg],
\end{equation}
where the coefficients are 
\begin{equation}\label{ZetaCoeffs}
\begin{split} 
  \mathcal{L}_{\partial_{i}\zeta\partial_{i}\zeta}&=2\Big[f-\frac{2}{a}\frac{d}{dt}(a\mathcal{M})\Big], \\
   \mathcal{L}_{\dot{\zeta}\dot{\zeta}}&=2\Big(c+2M_{2}^{4}-3H^{2}f-3H\dot{f} 
    \\
    &+3Hm_{3}^{3}-6H^{2}m_{4}^{2}\Big)\mathcal{D}^{2} +6(f+2m_{4}^{2}),
\end{split}
\end{equation}
with
\begin{equation}
    \begin{split} 
        \mathcal{M}&=\frac{\mathcal{D}}{2}(f+2\tilde{m}_{4}^{2}), \\
        \mathcal{D}&=\frac{2f+4m_{4}^{2}}{2H(f+2m_{4}^{2})+\dot{f}-m_{3}^{3}}.
    \end{split}
\end{equation}
The sound speed is computed in the EFT by taking the ratio
\begin{equation}\label{speed2}
    c_{s}^{2}=-\frac{\mathcal{L}_{\partial_{i}\zeta\partial_{i}\zeta}}{\mathcal{L}_{\dot{\zeta}\dot{\zeta}}}.
\end{equation}

The last step of the program is to translate the fundamental Lagrangian, (\ref{Sphi}), into the EFT language. For our purposes, we only need to study the two operators in the third line of (\ref{EFT}) for the case of minimal scalar-tensor coupling, $f(t)=1.$ First, the background functions are $\lambda(t)=\dot{H}+3H^{2}$, $c(t)=-\dot{H}$, and we set $m_{3}^{3}=m_{4}^{2}=0.$ Next, observe that the coefficient $M_{2}^{4}(t)$ is fixed in terms of $p(\varphi,X)$ in (\ref{Sphi}). This can be seen as follows: since the perturbation of the inverse metric is defined as $\delta g^{00}=g^{00}+1,$ expanding the Lagrangian (\ref{Sphi}) about a background kinetic term $X_{0}=\frac{1}{2}\dot{\varphi}^{2}$ and comparing to (\ref{EFT}) shows
\begin{equation}\label{M24}
    M_{2}^{4}(t)=\frac{1}{4}\dot{\varphi}(t)^{4}p_{XX},
\end{equation}
which we return to below. Finally, the coefficient $\tilde{m}_{4}^{2}$ of the $\mathcal{R}\delta g^{00}$ operator is taken to be a free parameter in a similar fashion to \cite{Cai1}, where it plays a role in the cosmological bounce scenario. One should note that this function drastically effects the shape of the sound speed (\ref{speed2}). As far as the authors know, theoretical constraints on $\tilde{m}_{4}^{2}$ are yet to be discovered. Nevertheless, there are strong phenomenological constraints on the function that will be discussed in our conclusion. 

We now have all the ingredients to compute the sound speed in the presence of the EFT operators and compare to the background calculation. The time crystal Lagrangian is given by \cite{Wilczek3}
\begin{equation}\label{Model}
    p(\varphi,X)=(3b\varphi^{2}-1)X+X^{2}-V(\varphi),
\end{equation}
with Higgs-like potential \footnote{The Lagrangian (\ref{Model}) with the potential (\ref{potential})  written in this form is strictly a toy model, where the three $O(1)$ parameters $\{b,d,\Lambda \}$ serve to illustrate its basic features. A realistic model necessitates reintroducing the scalar's mass on the quadratic term in the potential, the Planck mass, as well as a dimensionful coupling on the $X^{2}$ term. The $X^{2}$ coupling can be shown to set a UV-cutoff of the model. Further discussion of the model's regime of validity can be found in \cite{Wilczek3}.}
\begin{equation}\label{potential}
    V(\varphi)=\Lambda+\frac{1}{12d}-\frac{1}{2}\varphi^{2}+\frac{3d}{4}\varphi^{4}.
\end{equation}
The energy density $\varepsilon=2Xp_{X}-p$ becomes
\begin{equation}
    \varepsilon=(3b\varphi^{2}-1)X+3X^{2}+V(\varphi)
\end{equation}
and the sound speed (computed on the background using $c_{s}^{2}=\frac{p_{X}}{\varepsilon_{X}}$) is
\begin{equation}\label{speed1}
    c_{s}^{2}(\varphi,\dot{\varphi})=1-\frac{2\dot{\varphi}^{2}}{3\dot{\varphi}^{2}+3b\varphi^{2}-1}.
\end{equation}
Recalling $X=-\frac{1}{2}(\nabla_{\mu}\varphi)^{2}\sim\frac{1}{2}\dot{\varphi}^{2}$, we can note that the ghost singularity $\varepsilon_{X}=0$ and NEC violation $p_{X}=0$ boundaries can be expressed as ellipses in the $(\varphi,\dot{\varphi})$ phase space:
\begin{equation}
    \begin{split}
        \varepsilon_{X}=0 \ \ \Rightarrow & \ \  3\dot{\varphi}^{2}+3b\varphi^{2}=1, \\
        p_{X}=0 \ \ \Rightarrow  & \ \ \dot{\varphi}^{2}+3b\varphi^{2}=1.
    \end{split}
\end{equation}
Both curves are plotted on the phase portraits in FIG. 1 in red and green, respectfully. The sound speed is superluminal inside the ellipse $\varepsilon_{X}=0$, which is referred to as the graveyard or ghost region. The gradient instability $c_{s}^{2}<0$ occurs outside of the ghost region but inside the NEC violating phase. Any solution on the limit cycle attractor, shown in yellow in FIG. 1, passes through this region twice in each cycle. All (non-ghost) solutions will eventually be drawn into the attractor to become time crystals, having nonzero ground state motion despite the usual presence of Hubble friction (see (\ref{PhiEom})). Thus the EFT operators are essential if we wish to avoid the gradient instability. We would like to reiterate that the addition of the operators does not effect the background solutions for $\varphi$ or the scale factor $a(t)$. Instead, the operators act like a sort of regulator for the sound speed, as we will see below. 

The coefficients in (\ref{ZetaCoeffs}) become
\begin{equation}
    \begin{split}
        \mathcal{L}_{\partial_{i}\zeta\partial_{i}\zeta}&=2\Big[1-\frac{1}{a}\frac{d}{dt}\Big(\frac{a}{H}(1+2\tilde{m}_{4}^{2})\Big)\Big] \,,\\
        \mathcal{L}_{\dot{\zeta}\dot{\zeta}}&=\frac{2(2M_{2}^{4}+c)}{H^{2}} \,,
    \end{split}
\end{equation}
after making our simplifications to (\ref{EFT}), and the sound speed (\ref{speed2}) is given by
\bea\label{FinalSpeed}
  c_{s}^{2}&=&\frac{H^{2}}{\dot{\varphi}^{4}-\dot{H}}\Bigg[1-\frac{\ddot{a}}{aH^{2}}+2\tilde{m}_{4}^{2}\Big(2-\frac{\ddot{a}}{aH^{2}}\Big) \nonumber \\
  &+&\frac{2}{H}\frac{d}{dt}\tilde{m}_{4}^{2}\Bigg] \,,
\eea
where we have used (\ref{M24}) and that $p_{XX}=2$ for the model (\ref{Model}). Similar to the field, the sound speed oscillates with a period $\mathcal{T}$. The gradient instability occurs for a comparatively short time $\tau<\mathcal{T}.$ We therefore want to turn on $\tilde{m}_{4}^{2}(t)$ only from $t^{*}$ to $t^{*}+\tau$, where $c_{s}^{2}(t^{*})=0$ is the first zero crossing~\footnote{We can guarantee that $c_{s}^{2}$ is an even, perfectly periodic function if we choose one of the two initial conditions lying on the limit cycle where $\dot{\varphi}_{0}\approx0$ so that $c_{s}^{2}=1$ initially, given by (\ref{speed1}). Solutions that begin off the limit cycle are not perfectly periodic until they converge to the limit cycle attractor.}. This is easily accomplished using a Fourier series. As is clearly seen in the phase space portraits FIG. 1 (a) and (b), different choices for the parameters in the Lagrangian result in limit cycles with varied behaviors (\it c.f. \rm \cite{Damien1}). Consequently, the shape of the sound speed will depend on the parameter values, FIG. 1 (c) and (d). Two examples of $\tilde{m}_{4}^{2}$ resolving the gradient instabilities are included on the plot. Note that the sound speed is subluminal for all time (a desirable property, deeply related to the notion of UV completion \cite{Adams:2006sv}). While our choices for $\tilde{m}_{4}^{2}$ are not unique, these simple functions suffice to show that the model can be stabilized.


\begin{figure*}
\label{fig1}
\centering 
$\begin{array}{cc}
\subfigure[]{
\includegraphics[width=0.4 \textwidth,clip]{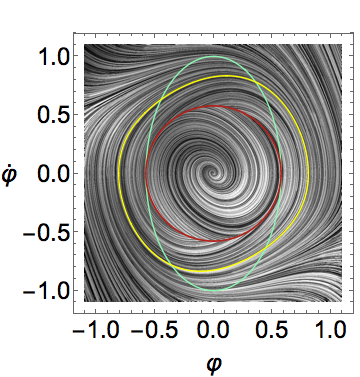}}
\subfigure[]{
\includegraphics[width=0.4 \textwidth,clip]{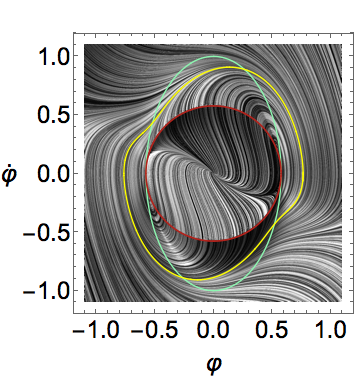}}
\end{array}$\\

$\begin{array}{cc}
\subfigure[]{
\includegraphics[width=0.4 \textwidth]{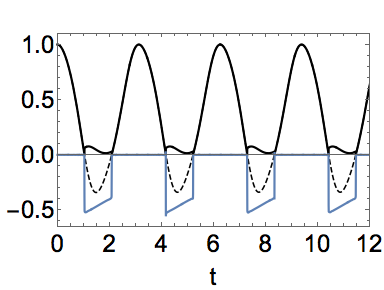}}
\subfigure[]{
\includegraphics[width=0.4 \textwidth,clip]{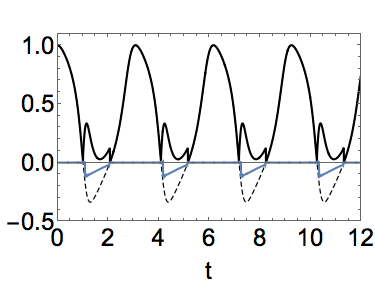}}
\end{array}$

\caption{\small{(a): Phase space for the parameter set $b=d=1,$ $\Lambda=0$ for the model (\ref{Model}). The limit cycle is depicted by the yellow curve. Inside the red curves are ghosts. The NEC is violated inside of the green ellipse. 
(b): Phase space for the parameter set $b=d=1,$ $\Lambda=1$.
(c): $c_s^{2}$ with $b=d=1,$ $\Lambda=0.$
(d): $c_s^{2}$ with $b=d=\Lambda=1.$ The sound speed computed with $\tilde{m}_{4}^{2}$ turned off, or equivalently by $c_{s}^{2}=\frac{p_{X}}{\varepsilon_{X}}$, is shown with the dashed lines, while the solid black curve is the stabilized sound speed. The corresponding  $\tilde{m}_{4}^{2}$ for each parameter set are shown in solid blue.}}
\end{figure*}


We have therefore successfully created stable cosmological time crystals. Regarding their numerous applications, we first note that they provide an inflationary scenario. Furthermore, both $H$ and $\dot{H}$ oscillate in time and $\dot{H}>0$ occurs periodically, which is referred to as \textit{super-inflation}. Time-crystals can therefore be used as a tool in studying this phenomenon in the early universe. As the universe expands, $\varphi$'s rapid oscillations will excite other fields and create particles. Interactions between $\varphi$ and newly created fields can 'break' the time crystal, providing a smooth exit from inflation. Moreover, and perhaps most interestingly, they may be natural candidates for playing a role in relaxing the cosmological constant via the mechanisms proposed in \cite{Creminelli3}. 

Finally, we wish to comment on the phenomenological aspects and constraints on the model by considering the temperature fluctuations in the Cosmic Microwave Background (CMB). The quantity of interest is the correlation function of the temperature fluctuations \cite{Mukhanov}, $C(\theta)\equiv\langle\frac{\delta T(\bm{l}_{1})}{T_{0}}\frac{\delta T(\bm{l}_{2})}{T_{0}}\rangle$ (the bracket denotes averaging over all directions $\bm{l}_{1}$, $\bm{l}_{2}$ where $\bm{l}_{1}\cdot\bm{l}_{2}=\cos\theta$). $C(\theta)$ can be decomposed as a sum over multipole moments using Legendre polynomials, $C(\theta)\propto\sum_{\ell}(2\ell+1)C_{\ell}P_{\ell}(\cos\theta)$, which can further be split into scalar and tensor contributions $C_{\ell}^{S}$ and $C_{\ell}^{T}$. For our model, the relative contribution of the tensor to scalar fluctuations at the quadrupole, $\ell=2,$ is approximated by
\begin{equation}\label{TSratio}
    \frac{C_{\ell=2}^{T}}{C_{\ell=2}^{S}}\simeq 10.4c_{s}\Big(1+p/\varepsilon\Big).
\end{equation}
The right hand side is to be evaluated when the perturbations responsible for the quadrupole fluctuation cross the Hubble scale during inflation. This ratio describes the amount primordial gravitational waves contribute to the $\ell=2$ component, and is crucially dependent on the sound speed. With a unit sound speed $c_{s}=1$ and assuming $1+p/\varepsilon\sim 10^{-2}$, gravitational waves contribute about 10\% to the quadrupole. Clearly, an oscillating sound speed can significantly suppress this ratio. Precision measurements of primordial gravitational waves, such as those expected from the European Space Agency's LISA project and other next-generation CMB experiments \footnote{For more information on the $LISA$ project see, e.g, \url{ http://sci.esa.int/lisa/}.}, will constrain not only the free parameters of the model (\ref{Model}), but also the form of the function $\tilde{m}_{4}^{2}$ through the sound speed (\ref{FinalSpeed}).

In summary, now that we have shown the time crystals can be stabilized, many facets of their applicability in cosmology merit further investigation.



We are delighted to thank Matthew Baumgart, Yong Cai and Frank Wilczek for helpful discussions.

\end{document}